\documentclass[%
aps, prb, twocolumn, showpacs, superscriptaddress, byrevtex,
 amsmath,amssymb,
]{revtex4-1}

\usepackage{graphicx}
\usepackage{dcolumn}
\usepackage{bm}

\begin{document}

\preprint{AIP/123-QED}

\title{Probing the spin states of three interacting electrons 
in quantum dots}

\author{A. Gamucci}

\author{V. Pellegrini}
\email{vittorio.pellegrini@nano.cnr.it.}

\author{A. Singha}

\affiliation{ 
CNR-NANO NEST and Scuola Normale Superiore, 
Piazza San Silvestro 12, 56127 Pisa, Italy
}

\author{A. Pinczuk}

\affiliation{Depts of Appl. Phys. \& Appl. Math. and of Physics, 
Columbia University, New York 10027, USA}

\author{L. N. Pfeiffer}

\author{K. W. West}

\affiliation{Dept of Electrical Engineering,
Princeton University, Princeton, and Bell Labs, Alcatel-Lucent, 
Murray Hill, New Jersey 07974, USA}

\author{M. Rontani}
\email{massimo.rontani@nano.cnr.it.}
\affiliation{%
CNR-NANO S3, Via Campi 213a, 41125 Modena, Italy
}%

\date{\today}

\begin{abstract}
We observe a low-lying sharp spin mode of three interacting electrons in 
an array of nanofabricated AlGaAs/GaAs quantum dots 
by means of resonant inelastic light scattering. 
The finding is enabled by a suppression of the inhomogeneous contribution to the excitation spectra 
obtained by reducing the number of optically-probed quantum dots.
Supported by configuration-interaction 
calculations we argue that the observed spin mode offers a direct probe of 
Stoner ferromagnetism in the simplest case of three interacting
spin one-half fermions.
\end{abstract}
\maketitle

%

Systems of three charged particles interacting by Coulomb forces 
are the building blocks of 
a large variety of correlated quantum phases. 
An example of current topical interest is represented
by the unusual fractional quantum Hall states with 
non-abelian excitations.\cite{moore,jain} 
On more general grounds, the interaction among at least three
particles is essential to account for 
Stoner ferromagnetism (SF), the paradigm of the tendency of  
itinerant electrons, like those responsible for conduction 
in metals, to align their spins at the expense of their kinetic 
energy.\cite{Stoner1933}

Indeed, the simplest finite-size version of SF displaying the crossover between 
normal (unpolarized) and spin-polarized degenerate (ferromagnetic) ground states
at a threshold interaction strength is offered by three interacting fermions of spin 
one-half in two dimensions, as: (i) in one dimension the ground state 
is never spin polarized for any interaction.\cite{LiebMattis} (ii) the 
two-body ground state is always a spin singlet.\cite{Lieb} 

Cold Fermi atoms confined in optical traps have been proposed 
as finite-size simulators of three-body interaction phenomena \cite{zoller} and SF physics. They are simple systems 
able to drive SF through the tunable short-range atom-atom 
interaction.\cite{Duine2005,Jo2009,Liu2010} However, this 
capability is limited by the losses due to three-body recombination. 

Here we realize and study a solid-state version of the quantum simulator 
of three interacting fermions based
on electrons confined in a semiconductor quantum dot (QD).
We show that unpolarized and ferromagnetic states can be probed by monitoring the low-lying neutral spin 
and charge excitations. We recall that such collective modes in semiconductor QDs 
can be accessed by resonant inelastic light scattering.\cite{Brocke2003,Garcia2005,Kalliakos2007,Kalliakos2008,Koppen2009,Singha2010} 
We argue that QD systems with three electrons constitute 
a versatile emulator of the physics of SF. 

So far studies on low-lying neutral spin modes in QDs were performed by optical methods 
on nanofabricated AlGaAs/GaAs QDs and in self-assembled InAs QDs. 
In these experiments, however, ensembles of many QDs were investigated 
owing to the very low signal-to-noise ratio. 
For example, arrays composed of 10$^4$  nanofabricated AlGaAs/GaAs QD 
replicas were studied by us by means of resonant inelastic light scattering.\cite{Garcia2005,Kalliakos2007,Kalliakos2008,Singha2010} 
In these systems, however, inhomogeneities in the electron number distribution 
among the QDs contribute to a significant broadening of the 
excitation peaks in the detected spectra, leading to a systematic 
uncertainty in the identification of specific contributions arising from different 
electron populations. Indeed, the emulation of SF requires access to
sharp collective excitations of three
electrons that are sufficiently isolated from excitations 
corresponding to other
electron number configurations.
This is achieved here in arrays of nanofabricated QDs which are diluted enough to 
suppress inhomogeneities linked to number fluctuations. 
We find that when the number of QD replicas probed in the light scattering experiments is 
decreased from $10^4$ to $10^3$, the inhomogeneities 
are largely suppressed, 
leading to a single and sharp spin mode in the excitation spectrum.

We argue in the following that this low-lying mode corresponds to a spin 
excitation of the three interacting electrons from the ground state with $M=1$, $S=1/2$ (being $M$ the 
total orbital angular momentum and $S$ the total spin of the system) to the 
excited state with $M=1$, $S=3/2$. The assignment  is corroborated 
by accurate full configuration interaction (CI) 
calculations.\cite{Rontani2004,Garcia2005,Rontani2005,Rontani2006,Kalliakos2008,Singha2010} 
The latter enable to link the measured excitation to the spin-polarized 
state with $M=0$, $S=3/2$ predicted by SF, which is an excited state at 
the actual value of the electron density.

\begin{figure}[ht]
\centering
\includegraphics[width=0.5\textwidth]{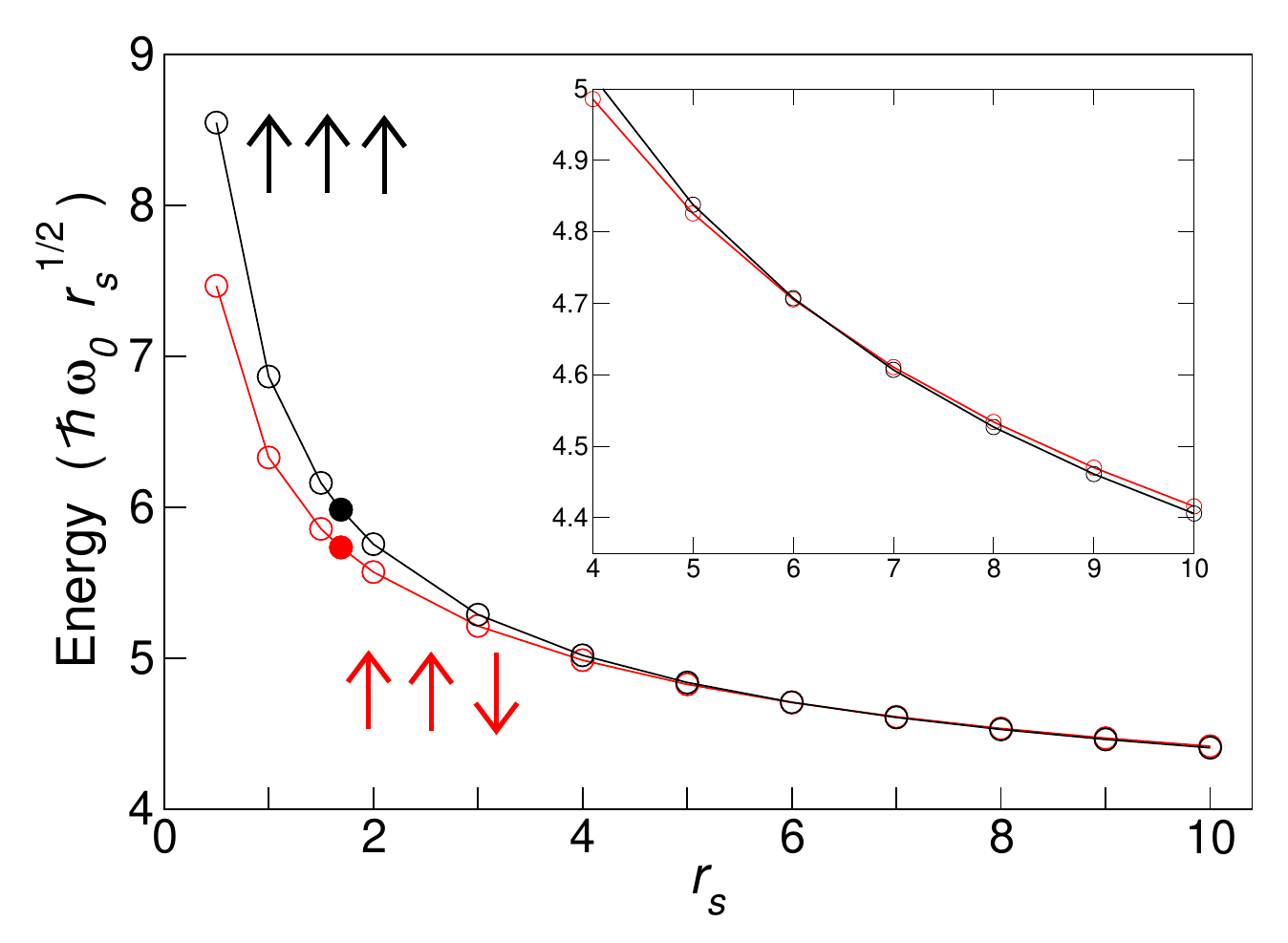}
\caption{(Color online) CI energies of the unpolarized (red [gray] circles) 
and spin-polarized (black circles) three-electron ground states vs $r_s$.
The filled circles point to the values inferred from the inelastic
light scattering measurement.
Inset: zoom around the crossover. Ten harmonic-oscillator shells  
were considered in the CI calculation. A square 
QW of width 25 nm and height 250 meV confines the motion in the out-of-plane
direction.
}
\label{fig_phase}
\end{figure}

The three-electron system exhibits a transition to a ferromagnetic
state as a function of the Wigner-Seitz parameter $r_s$,
which is the radius of the circle whose area is the average area per 
electron, in units of the Bohr radius.
By knowing both the energy of the transition and the value of the
energy spacing $\hbar\omega_0$ among the shells of the potential trap, 
as retrieved from the 
measurements, we can thus deduce the energy difference between the absolute 
ground state of the system and the fully spin-polarized ground state. 
As it is shown in Fig.~\ref{fig_phase}, this makes it possible to place 
our results (filled circles in Fig.~\ref{fig_phase})
on a phase diagram showing the path to the realization of SF.

We recall that the essence of SF is summarized by the `Stoner criterion'.\cite{Grosso2000}
According to that,
the ground state is ferromagnetic when $n(E_F) J > 1$,
with $n(E_F)$ being the density of states resolved at the Fermi energy
and $J$ being proportional to the exchange field that splits the energies of electrons
of opposite spin. Hence two parameters, namely $J$ and
$n(E_F)$, control the relevant physics of the bulk.

\begin{figure}[!ht]
\centering
\includegraphics[width=0.5\textwidth]{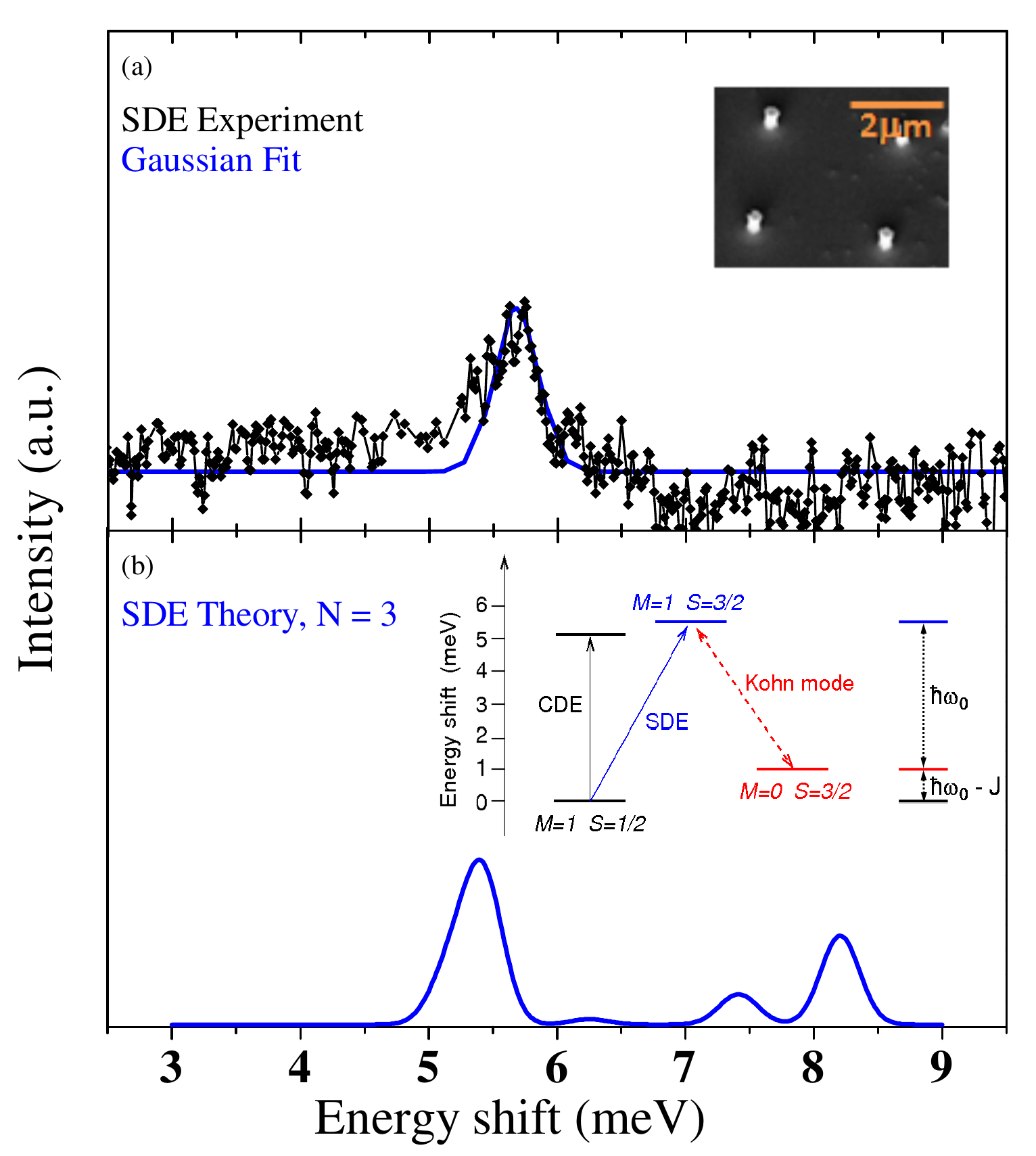}
\caption{(Color online) (a) Experimental resonant inelastic light scattering 
spectrum (black curve with points) 
and Gaussian fit (continuous blue [dark grey] curve) for the spin channel for 
the 1110-QD array. A scanning electron microscope image of the relative QD array is also shown. (b) Configuration-interaction calculations
of spin density excitation for $N=3$. The inset shows the energy diagram of relevant transitions.}
\label{fig_last}
\end{figure}

The finite-size analog of the criterion, derived in a simple Hartree-Fock 
picture, is $J / (\hbar\omega_0) > 1$, with $ 1 / (\hbar\omega_0)$ 
replacing the density of states $n(E_F)$. 
In our quantum simulator, the first parameter,
$\hbar\omega_0$, is obtained through the joint measure of both
the electron number $N$ in the QD and the pristine electron density $n$ 
of the quantum well from which QD arrays are nanofabricated.\cite{Garcia2005} 
For fixed $N$, $\hbar\omega_0$ is decreased 
($r_s$ is increased) by decreasing $n$.
The last parameter, $J$, is measured from the lowest spin excitation,
that is $2\hbar\omega_0 - J$
(cf.~inset of Fig.~\ref{fig_last}).

A series of QD arrays were fabricated on a 25~nm modulation doped 
GaAs/Al$_{0.1}$Ga$_{0.9}$As quantum well (QW) by means of state-of-the-art 
electron beam lithography and inductively coupled-plasma reactive 
ion etching. The electron density of the two-dimensional electron gas was 
$n=1.1\cdot$10$^{11}$~cm$^{-2}$, and the mobility 
2.7$\cdot$10$^{6}$~cm$^{2}$/Vs. Pillars with diameters 
of 320~nm and aspect ratio of $\sim$2/5 were produced 
in 0.1 mm-sized square arrays, in which the number of QD replicas was 
varied from 10000 down to $\approx1000$ to explore the impact of inhomogeneous broadening on the collective excitations. 
Resonant inelastic light scattering measurements were performed 
at $T=2$ K with a tunable ring-etalon Ti:Sa laser impinging with a spot 
size of 0.1 mm at normal incidence on the sample. The scattered signal from the QDs was collected by a triple grating 
spectrometer coupled to liquid N$_2$-cooled multichannel CCD. From previous experiments, it follows 
that a number of electrons ranging from 2 to 6 are expected to be 
confined in such QDs with a confinement energy of $\hbar\omega_0\approx 4$ meV.\cite{Garcia2005,Kalliakos2007,Kalliakos2008}

\begin{figure}[!ht]
\centering
\includegraphics[width=0.5\textwidth]{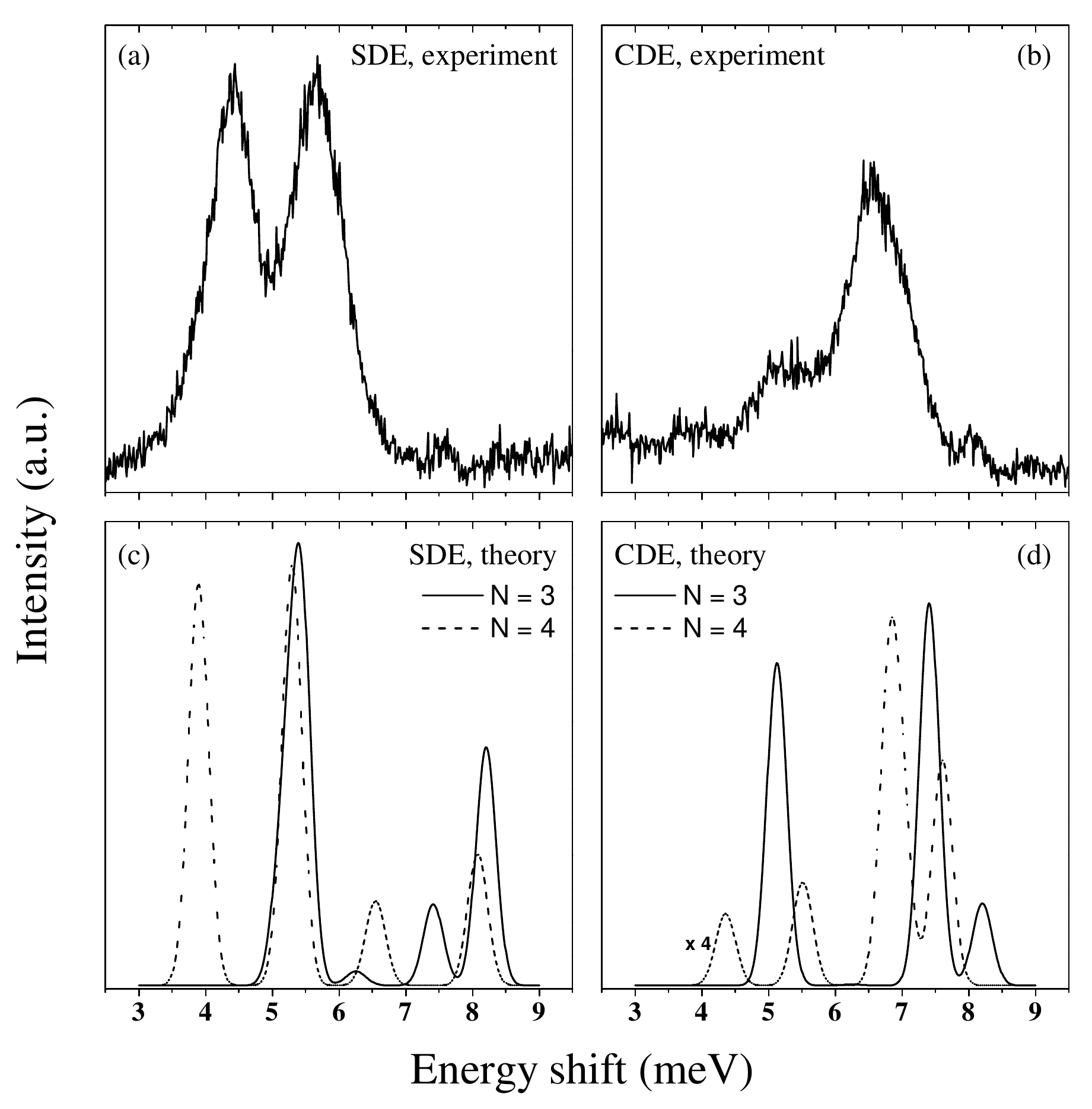}
\caption{Top panels: experimental spectra of spin (a) and charge (b) density 
excitations detected at $T=2$ K by means of depolarized and polarized 
resonant inelastic light scattering, respectively. 
Bottom panels: Corresponding CI spectra obtained for $N=3$ (solid line) and 
$N=4$ (dashed line).}
\label{fig_spectra10000}
\end{figure}

We focus first on the sample with $10000$ QD replicas.
Figures~\ref{fig_spectra10000}(a) and ~\ref{fig_spectra10000}(b) show 
the results of light scattering measurements of spin (SDE) and charge density excitations
(CDE), respectively, obtained with depolarized (perpendicular incident 
and scattered laser polarizations to detect the spin signal) and 
polarized (parallel incident and scattered laser polarizations to detect 
the charge signal) configurations. 
Two prominent peaks emerge in the spin channel at energies of about 4.4 meV 
and 5.6 meV, respectively, with the same intensity and full width at 
half maximum (FWHM) of approximately 1 meV. This value matches with the 
results of previous experiments, in which an average population of 
four electrons per dot was identified.\cite{Garcia2005,Kalliakos2008} 

\begin{figure}[!ht]
\centering
\includegraphics[width=0.5\textwidth]{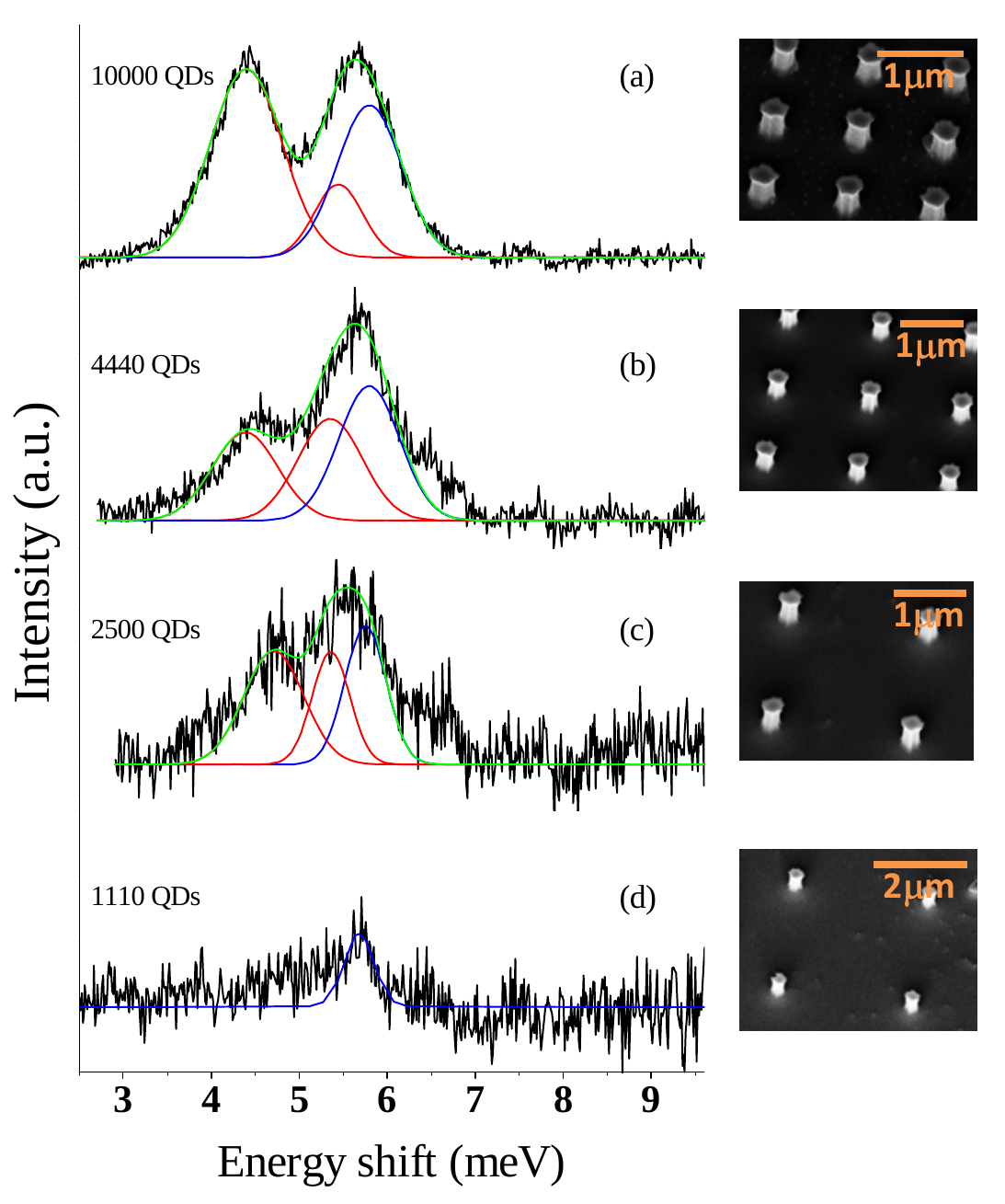}
\caption{(Color online) Resonant inelastic light scattering spectra of 
spin excitations obtained at $T=2$ K 
for arrays containing a varying number of QDs. From top to bottom, 
the number of QDs in the array decreases from 10000 to 1110 
(the inter-dot distance increases from 1 $\mu$m to 3 $\mu$m). 
Gaussian fits are superimposed to the experimental spectra in 
correspondence of three- and four-electron contributions 
(blue [dark grey] and red [gray] curves, respectively). 
The resulting global fit is shown as a green [light gray] curve on each 
spectrum. On the right hand side, a scanning electron microscope 
image of the studied QDs is shown, in correspondence to each case.}
\label{fig_trend}
\end{figure}

In order to single out the contributions linked to specific 
electron populations, we applied the full 
CI approach to retrieve the theoretical 
spectra linked to a series of electron populations, ranging from 
$N=2$ to $N=6$, with $\hbar\omega_0$ being fixed by
the experimental value of $n$. 
This analysis, which has spanned the wider expected range 
of confined electron numbers, enabled the identification of two prominent 
contributions to the experimental spectra arising 
from $N=3$ and $N=4$ electrons. 
Figures~\ref{fig_spectra10000}(c) and ~\ref{fig_spectra10000}(d) show 
the agreement with the experimental data obtained in the case of three 
and four electrons per QD. In both spin and charge channels, the interplay of the contributions from these two
electron configurations reflects in a wider peak at higher energy.
In the SDE spectrum, in particular,  there is a strong feature 
associated to the three-electron case at around 5.5 meV, which
overlaps with the contribution of the four-electron configuration. 
Higher-energy peaks resolved in the theoretical simulation
are not visible in the experimental spectra probably due to their weaker intensities.
The evolution of the SDE experimental spectra as the 
number of QDs in the illuminated array is progressively decreased is shown in Fig.~\ref{fig_trend}. 
These data confirm the interplay between these two different electron populations 
allowing to isolate the specific contribution of the QDs with three interacting electrons. 
The measurements carried out with varying the density of 
QDs allow us to confirm the broadening of the high-energy 
peak as mainly due to the simultaneous presence of 
contributions associated to a different number of electrons, namely 
$N=3$ and $N=4$ as modeled by the theoretical analysis. 
In Fig.~\ref{fig_trend}, Gaussian fits centered at Raman energies matching 
with the CI predictions [Fig.~\ref{fig_spectra10000}(c)] have been 
considered in order to globally reproduce the experimental data. 
It follows that: (i) the two slightly shifted peaks relative to $N=3$ 
(blue [dark gray] curve in Fig.~\ref{fig_trend}) and $N=4$ 
(red [gray] curve)
present at $\sim$~5.6~meV broaden the overall parent structure (green
[light gray] curve) to 
$\sim$ 1 meV FWHM for 10000-QD array [Fig.~\ref{fig_trend}(a)]. 
(ii) the progressive narrowing 
of the peaks due to the vanishing of the $N=4$ feature in the less dense 
arrays makes the linewidth of the high-energy peak decrease down to the 
limiting value of 0.4 meV FWHM for 1110 QDs [blue (dark gray) curve 
in Fig.~\ref{fig_trend}(d)]. 
We thus eventually managed to realize a system in which only the population 
of three electrons is present. 
The spectrum of spin excitation measured for the array of 1110 QDs 
is reproduced in Fig.~\ref{fig_last} together with the pertinent CI
prediction for $N=3$.  

By comparing the top and bottom panels of Fig.~\ref{fig_last} and 
Fig.~\ref{fig_trend}, it emerges that the $N=3$ contribution has been 
identified and selected out of a more complex background. 
The lowness of the signal, at the limit of the signal-to-noise detection 
threshold, implies that only the most intense peak, centered 
at $\sim$~5.5~meV, can be safely identified.

The transition energies involved in the measurement are indicated in the 
inset to Fig.~\ref{fig_last}. The retrieved monopole SDE links the 
ground state, characterized by $M=1$ and $S=1/2$, to an 
excited state with $M=1$, $S=3/2$. The final state of this transition 
is also accessible via a Kohn mode excitation from the
ferromagnetic ground state with $M=0$, $S=3/2$.
Remarkably, the two spin-polarized states with $M=0$ and $M=1$ differ
only in the center-of-mass motion, whose excitation---exactly of
energy $\hbar
\omega_0$---is insensitive to interactions. Therefore, with regards to
the non-trivial part of the wave function that is affected by
interactions, the final state of the observed SDE is a replica of the
ferromagnetic ground state.
This attribution is crucial to address the realization of a spin-polarized 
state relevant to the study of Stoner ferromagnetism.

In conclusion, we have identified the low-lying modes of three interacting electrons in a quantum dot 
by means of inelastic light scattering. By decreasing the number of optically-probed QDs we were able to 
suppress the inhomogeneities related to the simultaneous presence 
of different electron populations leading to a spin excitation spectrum dominated
by a single and sharp (FWHM of 0.4 meV) peak. By a detailed comparison
with calculations based on a  configuration interaction method we have
linked to observed peak to a transition from the $M=1$, $S=1/2$ ground state to the $M=1$, $S=3/2$ 
excited state of the three interacting electrons.These results mark a starting point for the 
simulation of Stoner ferromagnetism with a three electron quantum dot 
and for the investigation of the competition between spin-unpolarized $S=1/2$ and 
spin-polarized $S=3/2$ states.

We thank G. Goldoni and E. Molinari for stimulating discussions.
This work is supported by projects MIUR-PRIN no. 2008H9ZAZR,
Fondazione Cassa di Risparmio di Modena `COLDandFEW', 
CINECA-ISCRA no. HP10BIFGH8.

\nocite{*}

\begin{thebibliography}{40}
\bibitem{moore} G. Moore and N. Read, Nucl. Phys. B {\bf 360,} 362 (1991).
\bibitem{jain} A. Wojs, C. Toke, and J. Jain, Phys. Rev. Lett. {\bf 105,} 196801 (2010).
\bibitem{Stoner1933} E.~C.~Stoner, Phil.~Mag. \textbf{15,} 1018 (1933).
\bibitem{LiebMattis} 
E.~Lieb and D.~Mattis, Phys.~Rev. \textbf{125,} 164-172 (1962).
\bibitem{Lieb} D. C. Mattis, {\it The Theory of Magnetism} (Harper,
New York, 1965).
\bibitem{zoller} H. P. B\"uchler, A. Micheli, and  P. Zoller, Nature Physics {\bf 3,} 726 (2007). 
\bibitem{Duine2005}
R. A. Duine and A. H. MacDonald, Phys. Rev. Lett. {\bf 95,}
230403 (2005).
\bibitem{Jo2009} G.-B.~Jo, Y.-R.~Lee, J.-H. Choi, C.~A.~Christensen, T.~H.~Kim,
J.~H.~Thywissen, D.~E.~Pritchard, and W.~Ketterle, Science \textbf{325,} 
1521-1524 (2009).
\bibitem{Liu2010} X.-J.~Liu, H.~Hu, and P.~D.~Drummond, 
Phys.~Rev.~A \textbf{82,} 023619 (2010).
\bibitem{Brocke2003} T.~Brocke, M.-T.~Bootsmann, M.~Tews, B.~Wunsch, D.~Pfannkuche, Ch.~Heyn, W.~Hansen, D.~Heitmann, and C.~Sch\"uller, Phys.~Rev.~Lett. \textbf{91,} 257401 (2003).
\bibitem{Garcia2005} C.~Pascual~Garcia, V.~Pellegrini, A.~Pinczuk, M.~Rontani, G.~Goldoni, E.~Molinari, B.~S.~Dennis, L.~N.~Pfeiffer, and K.~W.~West, Phys.~Rev.~Lett. \textbf{95,} 266806 (2005).
\bibitem{Kalliakos2007} S.~Kalliakos, C.~P.~Garcia, V.~Pellegrini, M.~Zamfirescu, L.~Cavigli, M.~Gurioli, A.~Vinattieri, A.~Pinczuk, B.~S.~Dennis, L.~N.~Pfeiffer, and K.~W.~West, Appl.~Phys.~Lett. \textbf{90,} 181902 (2007).
\bibitem{Kalliakos2008} S.~Kalliakos, M.~Rontani, V.~Pellegrini, C.~P.~Garcia, A.~Pinczuk, G.~Goldoni, E.~Molinari, L.~N.~Pfeiffer, and K.~W.~West, Nature~Phys. \textbf{4,} 467-471 (2008).
\bibitem{Koppen2009} T.~K\"oppen, D.~Franz, A.~Schramm, Ch.~Heyn, D.~Heitmann, and T.~Kipp, Phys.~Rev.~Lett. \textbf{103,} 037402 (2009).
\bibitem{Singha2010} A.~Singha, V.~Pellegrini, A.~Pinczuk, L.~N.~Pfeiffer, K.~W.~West, and M.~Rontani,  Phys.~Rev.~Lett. \textbf{104,} 246802 (2010).
\bibitem{Grosso2000} G. Grosso and G. Pastori Parravicini, \emph{Solid State Physics}
(Academic Press, San Diego, 2000).
\bibitem{Rontani2004} M.~Rontani, S.~Amaha, K.~Muraki, F.~Manghi, E.~Molinari, S.~Tarucha, and D.~G.~Austing, Phys.~Rev.~B \textbf{69,} 085327 (2004).
\bibitem{Rontani2005} M.~Rontani and E.~Molinari, Phys.~Rev.~B \textbf{69,} 233106 (2005).
\bibitem{Rontani2006} M.~Rontani, C.~Cavazzoni, D.~Bellucci, and G.~Goldoni, J.~Chem.~Phys. \textbf{124,} 124102 (2006).

\end{thebibliography}

\end{document}